\documentclass{article}

\usepackage{arxiv}

\usepackage[utf8]{inputenc}
\usepackage[T1]{fontenc}
\usepackage{hyperref}
\usepackage{url}
\usepackage{booktabs}
\usepackage{amsfonts}
\usepackage{nicefrac}
\usepackage{microtype}
\usepackage{lipsum}
\usepackage{graphicx}
\usepackage{doi}
\usepackage{xcolor}
\usepackage{float}

\usepackage{svg}
\usepackage{tikz}
\usetikzlibrary{arrows.meta, positioning, shapes.geometric,
backgrounds, calc, arrows, automata, fit}
\usepackage{siunitx}
\usepackage{multirow}
\usepackage{layouts}
\usepackage{xspace}
\usepackage{amsthm}
\usepackage{amsmath}
\usepackage[numbers]{natbib}
\usepackage{cleveref}

\graphicspath{{figures/}}

\newcommand{\printtextwidth}{%
  \the\textwidth
}

\theoremstyle{plain}

\theoremstyle{remark}

\theoremstyle{definition}

\newcommand{\model}{%
  STAR-GO\xspace
}
\newcommand{\red}[1]{\textcolor{black}{#1}}
\usepackage{fancyhdr}
\title{STAR-GO: Improving Protein Function Prediction by Learning to Hierarchically Integrate Ontology-Informed Semantic Embeddings}

\author{
  \href{https://orcid.org/0009-0000-0657-7386}{\includegraphics[scale=0.06]{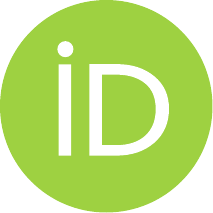}\hspace{1mm}Mehmet Efe Akca} \\
  \normalfont Department of Computer Engineering \\
  \normalfont Bogazici University \\
  \normalfont Istanbul, Turkiye \\
  \And
  \href{https://orcid.org/0000-0002-8684-2457}{\includegraphics[scale=0.06]{orcid.pdf}\hspace{1mm}Gökçe Uludoğan} \\
  \normalfont Department of Computer Engineering \\
  \normalfont Bogazici University \\
  \normalfont Istanbul, Turkiye \\
  \AND
  \href{https://orcid.org/0000-0001-8376-1056}{\includegraphics[scale=0.06]{orcid.pdf}\hspace{1mm}Arzucan Özgür}\thanks{Corresponding author: \href{mailto:arzucan.ozgur@bogazici.edu.tr}{arzucan.ozgur@bogazici.edu.tr}} \\
  \normalfont Department of Computer Engineering \\
  \normalfont Bogazici University \\
  \normalfont Istanbul, Turkiye \\
  \And
  \href{https://orcid.org/0000-0003-4765-2615}{\includegraphics[scale=0.06]{orcid.pdf}\hspace{1mm}İnci M. Baytaş}\thanks{Corresponding author: \href{mailto:inci.baytas@bogazici.edu.tr}{inci.baytas@bogazici.edu.tr}} \\
  \normalfont Department of Computer Engineering \\
  \normalfont Bogazici University \\
  \normalfont Istanbul, Turkiye \\
}

\date{}

\begin{document}
\sloppy

\maketitle
\begingroup
\renewcommand\thefootnote{}
\footnotetext{
This article has been accepted for publication in Bioinformatics, Published by Oxford University Press.
}
\endgroup

\begin{abstract}
  \textbf{Motivation:} 
  Accurate prediction of protein function is essential for elucidating molecular mechanisms and advancing biological and therapeutic discovery. Yet experimental annotation lags far behind the rapid growth of protein sequence data. Computational approaches address this gap by associating proteins with Gene Ontology (GO) terms, which encode functional knowledge through hierarchical relations and textual definitions. However, existing models often emphasize one modality over the other, limiting their ability to generalize, particularly to unseen or newly introduced GO terms that frequently arise as the ontology evolves, and making the previously trained models outdated. \\
  \textbf{Results:} 
  We present \model, a Transformer-based framework that jointly models the semantic and structural characteristics of GO terms to enhance zero-shot protein function prediction. STAR-GO integrates textual definitions with ontology graph structure to learn unified GO representations, which are processed in hierarchical order to propagate information from general to specific terms. These representations are then aligned with protein sequence embeddings to capture sequence–function relationships. \model achieves state-of-the-art performance and superior zero-shot generalization, demonstrating the utility of integrating semantics and structure for robust and adaptable protein function prediction. \\ 
  \textbf{Availability:} Code and pre-trained models are available at
  \url{https://github.com/boun-tabi-lifelu/stargo} \\ \red{(\url{https://doi.org/10.5281/zenodo.18643082})} 
\end{abstract}


\section{Introduction}
Proteins are essential macromolecules that perform diverse cellular functions, including catalyzing biochemical reactions, transmitting signals, and providing structural support. Understanding protein function, which is fundamental to biological research and therapeutic development, is a resource-intensive and time-consuming process. An extensive protein database, UniProt, contains over 190 million sequences of which only less than 1\% have experimentally determined functional annotations~\citep{theuniprotconsortiumUniProtUniversalProtein2021}. Consequently, computational function prediction methods have become crucial for addressing this functional annotation gap.

The annotations of protein functions are standardized through Gene Ontology (GO) \citep{consortiumGeneOntologyTool2000,geneontologyconsortiumGeneOntologyKnowledgebase2023}, a hierarchical classification scheme of three domains: molecular function (MF), biological process (BP), and cellular component (CC). GO contains a variety of information through its graph structure (e.g., ancestor-descendant relationships, part-of relationships, and regulation relationships), logical axioms, and textual descriptions (class labels and definitions).
The rich GO structure presents both opportunities and challenges for computational approaches.

Protein function prediction is commonly posed as multi-label classification over the GO terms. While early studies transferred annotations from homologous proteins~\citep{altschulBasicLocalAlignment1990}, recent methods design deep function prediction models, integrating protein sequences~\citep{kulmanovDeepGOPlusImprovedProtein2020, caoTALETransformerbasedProtein2021, sandersonProteInferDeepNeural2023}, structures~\citep{gligorijevicStructurebasedProteinFunction2021}, their interactions~\citep{you2021deepgraphgo}, and literature~\citep{yan2024goretriever}. Despite their versatility, most deep protein function prediction models treat GO terms as independent labels, neglecting their directed acyclic graph (DAG) nature, which is characterized by rich axioms and curated term descriptions. However, these hierarchical and semantic features can be used to relate GO terms with proteins, thereby improving the generalization capability to novel functions, particularly in zero-shot prediction, where the goal is to infer functions linked to GO terms unseen during training or newly introduced in future ontology releases. Therefore, recent studies propose embedding methods that map ontology entities into continuous vector spaces, preserving their structure and relationships to incorporate GO characteristics into the learning problem. 




General-purpose ontology embedding techniques such as OWL2Vec and EL Embeddings have been proposed and applied to GO. OWL2Vec integrates graph topology via random walks while combining logical axioms and lexical information~\citep{chenOWL2VecEmbeddingOWL2021}, whereas EL Embeddings ~\citep{Kulmanov_2019} represent ontological relations geometrically in a continuous space. In addition, GO-specific ontology embedding techniques have also been introduced. anc2vec~\citep{ederaAnc2vecEmbeddingGene2022} is a GO-specific ontology embedding technique, where the ontological uniqueness, ancestor hierarchy, and sub-ontology memberships are embedded via an autoencoder setting. GT2Vec \citep{zhaoLearningRepresentationsGene2022} also employs an autoencoder to generate semantic GO embeddings by fine-tuning BioBERT~\citep{leeBioBERTPretrainedBiomedical2020} while preserving graph structure. 

In line with advances in GO embedding techniques, protein function prediction models that incorporate GO embeddings into their architectures have also been recently proposed. DeepGOZero \citep{kulmanov2022deepgozero} and its successor DeepGO-SE \citep{kulmanovProteinFunctionPrediction2024} employ EL Embeddings, representing GO terms as geometric objects (n-dimensional balls) constrained by logical axioms. This formulation encodes the ontology's formal relationships, enabling zero-shot prediction for unseen GO terms. DeepGO-SE further introduces approximate semantic entailment by ensembling models built on pretrained protein language models \citep{rives2019biological}, yet its representations remain fundamentally axiom-based. 
PFresGO~\citep{panPFresGOAttentionMechanismbased2023} instead employs anc2vec~\citep{ederaAnc2vecEmbeddingGene2022} embeddings and integrates them with protein representations via cross-attention. While it captures hierarchical and membership relations, anc2Vec encodes GO terms as one-hot vectors, which limits its zero-shot capability.  \red{TransFew~\citep{transfew} combines BioBERT-derived textual embeddings with GCN-encoded hierarchical relationships and fuses them with protein representations via cross-attention, targeting rare GO term prediction. Although these studies benefit from incorporating ontological structure into protein function prediction, none jointly optimizes semantic and structural GO embeddings within a hierarchical decoding framework that captures functional dependencies across ontology levels.}

This study introduces \model, a Transformer-based framework that jointly embeds the structural and semantic characteristics of GO terms to enhance zero-shot protein function prediction. \model integrates hierarchical relations and textual definitions of GO terms, aligning ontology-informed embeddings with protein sequence representations to enable prediction of unseen functions. The framework has two main contributions: (i) GO term embeddings derived from a language model are refined via a structure-recovering autoencoder trained with multi-task supervision, preserving both semantic similarity and hierarchical dependencies for zero-shot inference without retraining; (ii) these enriched embeddings are incorporated into an encoder–decoder transformer, where GO terms are decoded in topological order using causal self-attention and linked to protein embeddings through cross-attention. The hierarchical decoding mechanism propagates information from general ancestors to specific child terms, capturing functional dependencies across levels. Evaluations under standard and zero-shot settings show that \model is competitive with state-of-the-art methods and demonstrates superior zero-shot generalization.

\section{Methods}

\subsection{Dataset}

We train and evaluate \model on a dataset curated from DeepFRI~\citep{gligorijevicStructurebasedProteinFunction2021}, which is commonly used in the protein function prediction literature. The dataset comprises $36,641$
protein sequences with coverage of $2,752$ GO terms, including the following subontology terms: $489$ Molecular Function,
$1,943$ Biological Process, and $320$ Cellular Component, where each GO term is linked to at least $50$ non-redundant Protein Data Bank
(PDB) chains.

The dataset is split into training (80\%; 29,902 sequences), validation (10\%; 3,323 sequences), and test (10\%; 3,416 sequences) sets (see Supplementary Table 1 for details). 
The test set consists only of proteins with at
least one experimentally validated functional annotation in each of the three GO
categories, with protein chain lengths limited to 1,000 residues. To
ensure non-redundancy
between the training and test sets, cd-hit \citep{10.1093/bioinformatics/btl158}
was applied using
95\% sequence identity threshold.

For zero-shot evaluation, we follow DeepGOZero's protocol~\citep{kulmanov2022deepgozero}, using the same similarity-based split of the UniProt/Swiss-Prot Knowledgebase (version 2021\_04, on September 29, 2021; see Supplementary Table 2).
We hold out the same 16 GO terms as DeepGOZero (5 MF, 7 BP, 4 CC), which are selected from classes with $>100$ annotations. We remove all protein-class associations for these terms from the training set before applying the true path rule, which propagates annotations of GO terms to their ancestors, ensuring hierarchical consistency.

We use the basic version of GO with the true path rule applied for training and evaluations, specifically the June 2020 release\footnote{\url{https://release.geneontology.org/2020-06-01/ontology/go-basic.obo}} in the standard setting and the November 2021 release\footnote{\url{https://release.geneontology.org/2021-11-16/ontology/go-basic.obo}} in the zero-shot setting.

\begin{figure*}[t]
  \centering
  \includegraphics[width=0.85\textwidth]{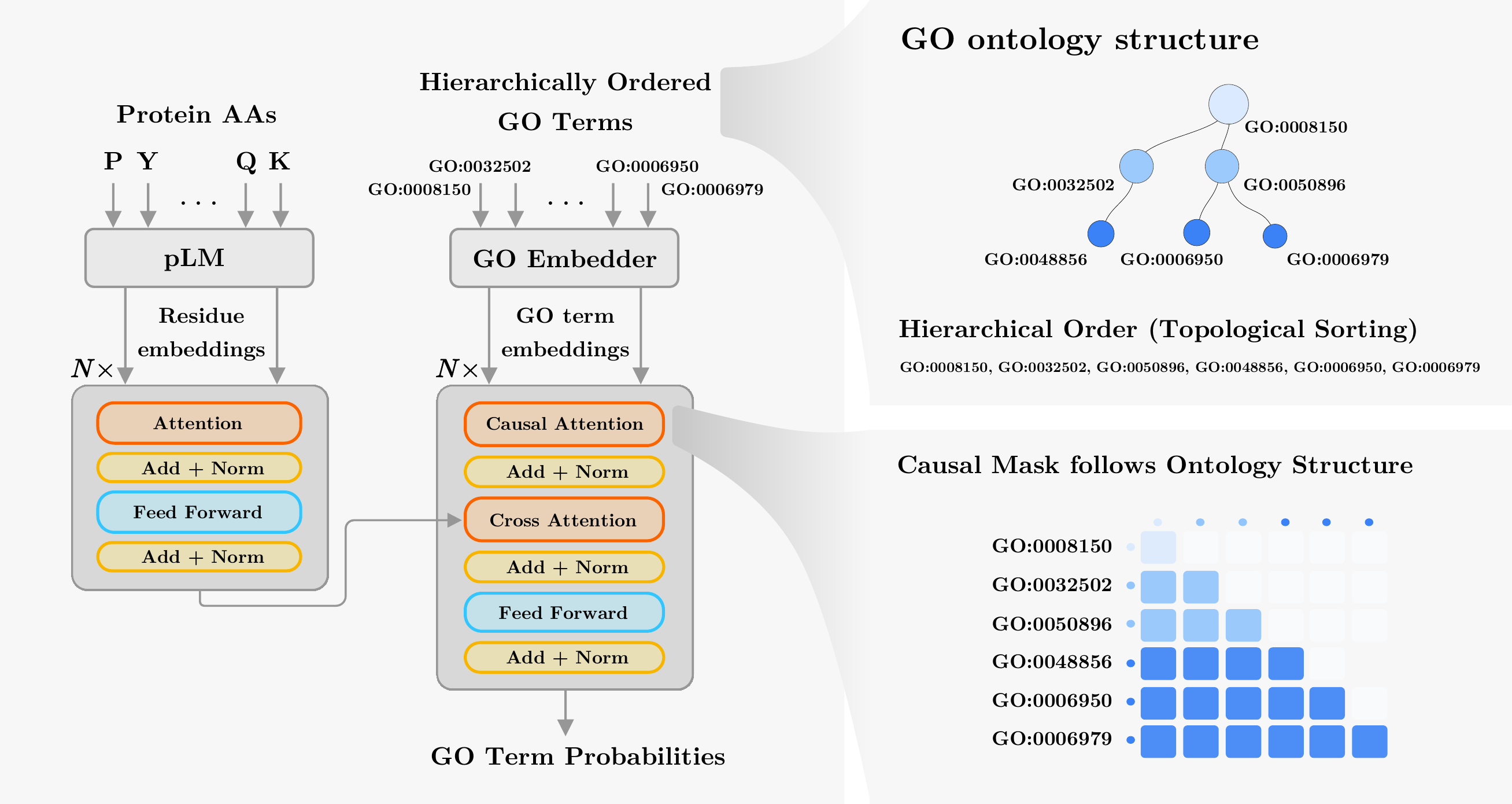}
  \caption{ Overview of the proposed STAR-GO architecture.
  The GO embedding module derives sentence-level representations of term descriptions (e.g., using SBERT-BioBERT) and projects them into a latent space that reflects the ontology's hierarchical structure.
  GO embeddings are hierarchically ordered, allowing the decoder to attend to ancestor terms and model dependencies across ontology levels.}
  \label{fig:model-architecture}
\end{figure*}

\subsection{\model}

\model is a Transformer-based architecture that integrates protein sequence representations with a novel GO embedding framework for protein function prediction. 
The GO embedder combines the semantic and structural information of GO terms by encoding term descriptions with their ontology relations.
The sentence-level embeddings of GO terms, extracted using a sentence encoder, are mapped into the latent space of an autoencoder jointly trained for predicting the GO term's aspect (i.e., MFO, BPO, CCO), its ancestor terms, and its identity. By jointly modeling semantic content and ontology hierarchy, this design enables zero-shot generalization to unseen GO terms.

\model adopts an \red{encoder-decoder} architecture, given in Figure \ref{fig:model-architecture}, where protein and GO embeddings are fused within a unified Transformer framework. 
The encoder takes protein embeddings obtained from a pretrained language model, while the decoder predicts GO term probabilities based on the fusion of the protein and GO representations via cross-attention. To ensure consistency with the hierarchical structure of the GO ontology, the decoder processes the GO embeddings arranged in a topological order from ancestors to descendants. Consequently, the proposed hierarchical decoding framework captures the dependencies across multiple functional levels.

\subsection{GO Embeddings}

GO is a structured vocabulary for protein functions, forming a directed acyclic graph (DAG), where nodes are terms and edges denote semantic relations \citep{consortiumGeneOntologyTool2000}.
The proposed embedding module aims to learn GO term embeddings by capturing the semantic meaning of term definitions and the structural relationships defined by the GO that are valuable for protein function prediction.
We obtain sentence embeddings for each GO term definition using SBERT-BioBERT~\citep{leeBioBERTPretrainedBiomedical2020}, pretrained with biomedical text. We apply mean pooling to token representations from its last layer to obtain GO term textual embeddings. The frozen embeddings serve as inputs to a trainable autoencoder that injects ontology structure via multi-task supervision~\citep{ederaAnc2vecEmbeddingGene2022}. 
Unlike anc2Vec, our autoencoder learns the hierarchical structure of GO terms from their sentence embeddings, which encode the semantic similarity between the function descriptions, rather than relying on one-hot representations.
The proposed encoder applies a linear projection, $\mathbf{z}_{t}= \mathbf{W}_{\text{emb}}\mathbf{x}_{t}$,
where $\mathbf{x}_{t}\in \mathbb{R}^{d_t}$ is the embedding of term $t$, and $\mathbf{W}_{\text{emb}}\in
\mathbb{R}^{d_{z}\times d_{t}}$ is the projection matrix.
The decoder learns to map $\mathbf{z}_{t}$, $\hat{y}_{\text{anc}}  =
  \mathbf{\phi}(\mathbf{W}_{\text{anc}}\mathbf{z}_{t}+
  \mathbf{b}_{\text{anc}}), \hat{y}_{\text{sub}}  =
  \mathbf{\phi}(\mathbf{W}_{\text{sub}}\mathbf{z}_{t}+
  \mathbf{b}_{\text{sub}}), \hat{y}_{\text{id}}   =
  \mathbf{\phi}(\mathbf{W}_{\text{id}}\mathbf{z}_{t}+
  \mathbf{b}_{\text{id}})$, using separate layers with softmax, $\phi\left(\cdot\right)$, to recover the term's ancestor set, its subontology, and its term identity with a combination of binary cross-entropy objectives:
\begin{align}
  \label{eq:BCE}\mathcal{L}_{\text{BCE}} & =
  -\alpha_{\text{anc}}\sum_{i=1}^{N_{\text{terms}}}y_{\text{anc}}^{i}\log
  \left(\hat{y}_{\text{anc}}^{i}\right)
  \\
  &
  -\alpha_{\text{sub}}\sum_{i=1}^{3}y_{\text{sub}}^{i}\log\left(\hat{y}_{\text{sub}}^{i}\right)-\alpha_{\text{id}}\sum_{i=1}^{N_{\text{terms}}}x_{i}\log
  \left(\hat{y}_{\text{id}}^{i}\right) \nonumber
\end{align}
where $\alpha_{\text{anc}}$, $\alpha_{\text{sub}}$, and $\alpha_{\text{id}}$ are the loss weights balancing ancestor, subontology, and term ID loss terms, respectively, $N_{\text{terms}}$ is the number of terms, and
$\mathbf{y}_{\text{anc}}\in \{0,1\}^{N_{\text{terms}}}$ denotes the ground truth
ancestor terms of $\mathbf{x}$, and $\mathbf{y}_{\text{sub}}\in
\{0,1\}^{3}$ is the
ground truth encoding of the subontology
terms. 

The anc2Vec~\citep{ederaAnc2vecEmbeddingGene2022} represents each GO term using one-hot vectors, where nonidentical terms are orthogonal and share limited similarity. In contrast, \model utilizes semantic embeddings of term descriptions, thereby it preserves relationships beyond the ontology graph. For example, Molecular Function terms GO:0005524 (ATP binding) and GO:0016887 (ATP hydrolysis activity) are functionally related since ATP hydrolysis inherently involves ATP binding. However, they share only the root ancestor molecular\_function in the ontology, illustrating the gap between semantics and graph structure. The cosine similarity between the one-hot vectors of these terms is $0.48$. With the anc2Vec embeddings, the similarity decreases to $0.29$, resulting in a limited structural overlap.
However, the similarities with our description-based embeddings are $0.69$ and $0.68$ before and after projection, respectively, indicating that semantic proximity is preserved. Finally, unlike anc2Vec, which assigns equal weights to each objective, we tune $\alpha_{\text{anc}}, \alpha_{\text{sub}}, \alpha_{\text{id}}$ to stabilize training and balance the influence of semantic and structural components.

\subsection{Transformer module}

The proposed module learns to associate protein sequences with GO terms by jointly processing residue-level embeddings representing protein sequences and term-level embeddings representing GO semantics and hierarchy. The architecture follows an \red{encoder-decoder} design where 
cross-attention combines protein and GO representations.
The encoder consists of $N_{\text{enc}}$ self-attention layers that contextualize residue embeddings, capturing dependencies along the sequence.
The decoder comprises $N_{\text{dec}}$ layers that combine self-attention and cross-attention operations.
The decoder's self-attention operates over GO terms with a causal mask that respects their topological order, allowing information to flow from general to specific functions per the GO hierarchy.
Cross-attention layers link the GO term representations with the residue embeddings, allowing each term to attend to the most informative residues relevant to its semantics. \red{We adopt cross-attention as the fusion mechanism because it has been successfully applied in prior protein function prediction models, including PFresGO and TransFew, to integrate protein and GO term representations.}
The model is trained for the supervised function prediction task, encouraging the model to capture task-relevant associations between residue-level and GO-level representations. Our implementation builds on HuggingFace Transformers~\citep{DBLP:journals/corr/abs-1910-03771} encoder and decoder layers. For a formal definition of the architecture, please see Supplementary Materials.

\textbf{Input and Projection.}
Each protein sequence is represented as residue-level embeddings. We extract these embeddings from the final hidden layer of a pretrained protein language model, \textbf{ProtT5}~\citep{ProtTransUnderstandingLanguage}, which is state-of-the-art across diverse protein prediction tasks.
For a sequence of length $L$, the residue embeddings, denoted as $\mathbf{X}\in\mathbb{R}^{L\times d_{\text{seq}}}$, are projected into the encoder's latent space:

\begin{equation}
\mathbf{H}^{(0)}_{\text{enc}} = \mathbf{X}\mathbf{W}_{\text{seq}} + \mathbf{1}\mathbf{b}_{\text{seq}}^{\!\top} \in\mathbb{R}^{L\times d}.
\end{equation}

\noindent \textbf{Self-Attention Layers.}
The encoder refines the projected residue embeddings through $N_{\text{enc}}$ stacked self-attention layers, each comprising multi-head attention~\citep{vaswaniAttentionAllYou2023}, a position-wise feed-forward network, and layer normalization with residual connections. After $N_{\text{enc}}$ layers, the encoder outputs refined residue embeddings
$\mathbf{H}^{(N_{\text{enc}})}_{\text{enc}}$.

\subsubsection{Decoder}
\noindent \textbf{Input Representation and Ordering.}
Each GO term within a subontology is represented by the embeddings obtained from the GO embedding module, $\mathbf{E}\in\mathbb{R}^{T\times d_{\text{go}}}$, where $T$ denotes the number of GO terms. 
To preserve the hierarchical dependencies among GO terms, they are arranged in a topological order 
$(\mathrm{GO}_{\pi(1)}, \ldots, \mathrm{GO}_{\pi(T)})$, 
where $\pi$ defines a breadth-first traversal from root to leaves. 
This ordering enables information to propagate along parent-child relationships during decoding.
The GO embeddings are then projected into the decoder's latent space, producing the initial representation $\mathbf{H}^{(0)}_{\text{dec}}$:
\begin{equation}
\mathbf{H}^{(0)}_{\text{dec}} = \mathbf{E}\mathbf{W}_{\text{go}} + \mathbf{1}\mathbf{b}_{\text{go}}^{\!\top} \in \mathbb{R}^{T\times d}.
\end{equation}

\noindent \textbf{Self-Attention and Cross-Attention.}
The decoder consists of $N_{\text{dec}}$ layers, each containing causal self-attention, cross-attention, and a feed-forward sub-layer with residual connections and layer normalization.
Causal self-attention is applied over GO terms using a lower-triangular mask $\tilde{\mathbf{M}}_{\text{dec}}$, \red{ensuring that each term attends only to terms preceding it in topological order, providing an inductive bias aligning with GO hierarchy}.
Subsequently, cross-attention incorporates the encoded protein representations $\mathbf{H}^{(N_{\text{enc}})}_{\text{enc}}$, allowing GO terms to attend to residue-level features relevant to their semantics. Thus, each GO term representation aggregates information from both its ancestors and the relevant residues of the protein.

\subsubsection{Prediction Head and Objective}
The final decoder output is mapped through a two-layer feed-forward projection with GELU activation and sigmoid to get the predicted probability that the protein is associated with each GO term $t$. The model is trained using binary cross-entropy.
During training, only the projection layers, Transformer blocks, and prediction head are optimized,
while the pretrained protein and GO embedding modules remain frozen.

\section{Results}
We compared \model with state-of-the-art models to evaluate its protein function prediction performance and generalization to unseen GO terms. We further examined the contribution of \model components with ablation studies.
Performance was evaluated using $F_{max}$ for protein-centric accuracy and Macro AUPR and AUC for term-centric generalization across GO terms. Implementation details, baselines, and evaluation metrics are provided in the Supplementary Materials.

\subsection{\model achieves competitive performance} 

We compared \model with state-of-the-art baselines, including PFresGO, DeepFRI, TALE+, DeepGOZero, DeepGO-SE, and \red{TransFew}, using $F_{\text{max}}$, Macro AUPR, and AUC metrics (Table~\ref{tab:baseline-comparison}).
Across all ontologies, \model exhibits competitive performance, matching or exceeding most baselines.
In BP, DeepGO-SE attains the highest overall $F_{\text{max}}$ and Macro AUPR, while \model achieves the best AUC (0.989).
Within CC ontology, \model surpasses all baselines in Macro AUPR and AUC while maintaining comparable $F_{\text{max}}$ to PFresGO.  For MF, \model performs on par with DeepGO-SE, achieving close $F_{\text{max}}$ and AUC values.
\model consistently ranks among the top-performing methods and achieves the strongest AUC values across all ontologies, indicating strong term-level discriminability.

\begin{table}[!htp]
  \begin{center}
    \begin{minipage}{\columnwidth}
      \caption{Comparison of STAR-GO with state-of-the-art. STAR-GO performs competitively across all ontologies. Notably, it obtains the highest AUC scores, demonstrating its term-level discriminability.}
      \label{tab:baseline-comparison}

      \tabcolsep=0pt
      \begin{tabular*}{\textwidth}{@{\extracolsep{\fill}}llccc@{\extracolsep{\fill}}}
        \toprule Ontology            & Method     & $F_{\text{max}}$
        & Macro AUPR     & AUC            \\
        \midrule \multirow{2}{*}{BP} & PFresGO    & 0.568          &
        0.293          & 0.839          \\
        & DeepFRI    & 0.540          & 0.261          & 0.858          \\
        & TALE+      & 0.554          & 0.302          & 0.811          \\
        & DeepGOZero & 0.565 & 0.294 & 0.768          \\
        & DeepGO-SE & \textbf{0.574} & \textbf{0.325} & 0.968          \\
        & \red{TransFew}   & \red{0.559}          & \red{0.254}          & \red{0.847}          \\
        & STAR-GO    & 0.548          & 0.288          & \textbf{0.989} \\
        \midrule \multirow{2}{*}{CC} & PFresGO    & \textbf{0.674} &
        0.361          & 0.884          \\
        & DeepFRI    & 0.613          & 0.274          & 0.884          \\
        & TALE+      & 0.610          & 0.325          & 0.849          \\
        & DeepGOZero & 0.534          & 0.315          & 0.738          \\
        & DeepGO-SE & 0.638          & 0.352          & 0.974          \\
        & \red{TransFew}   & \red{0.650}          & \red{0.347}          & \red{0.890}          \\
        & STAR-GO    & 0.659          & \textbf{0.379} & \textbf{0.988} \\
        \midrule \multirow{2}{*}{MF} & PFresGO    & 0.691          &
        0.602          & 0.924          \\
        & DeepFRI    & 0.625          & 0.494          & 0.915          \\
        & TALE+      & 0.662          & 0.564          & 0.884          \\
        & DeepGOZero & 0.719           & 0.614          & 0.893 \\
        & DeepGO-SE & \textbf{0.722} & \textbf{0.648}          & 0.991          \\
        & \red{TransFew}   & \red{0.695} & \red{0.593} & \red{0.946} \\
        & STAR-GO    & 0.719 & 0.620 & \textbf{0.995} \\
        \bottomrule
      \end{tabular*}
    \end{minipage}
  \end{center}
\end{table}

\subsection{\model generalizes to unseen GO terms}

\begin{table*}[!htp]
  \small
  \caption{
  Zero-shot performance of STAR-GO with unseen GO term AUCs. The zero-shot columns denote that models were optimized on an ablated dataset excluding the reported GO terms from all protein–term associations. Supervised columns reports the results for the corresponding protein-term pairs included during training. STAR variants use different GO embeddings: $S$ (structural, anc2Vec), $T$ (textual, SBERT-BioBERT), and $ST$ (combined, proposed in this work). The best AUC within each section is shown in \textbf{bold}. 
  }
  \label{tab:zero-shot-comparison}
  \centering
  \setlength{\tabcolsep}{4pt}
  \renewcommand{\arraystretch}{0.96}
  \resizebox{\textwidth}{!}{%
  \begin{tabular}{llrrrrrrrrr}
    \toprule
    &            & \multicolumn{6}{c}{Zero-shot} &
    \multicolumn{3}{c}{Supervised} \\

    \cmidrule(r){3-8} \cmidrule(l){9-11}
    Ontology & Term & STAR$_{ST}$ & STAR$_{T}$ & STAR$_{S}$ & DeepGOZero & DeepGO-SE & \red{TransFew} & STAR$_{ST}$ & DeepGOZero & DeepGO-SE \\
    \midrule 
    \multirow{5}{*}{MF} 
    & GO:0001227 & 0.891 & \textbf{0.944} & 0.557 & 0.257 & 0.759 & \red{0.395}& \textbf{0.955} & 0.932 & 0.952 \\
    & GO:0001228 & 0.938 & \textbf{0.949} & 0.508 & 0.574 & 0.791 & \red{0.533}& \textbf{0.961} & 0.948 & \textbf{0.961} \\
    & GO:0003735 & \textbf{0.923} & 0.915 & 0.301 & 0.400 & 0.066 & \red{0.459}& 0.994 & 0.940 & \textbf{0.995} \\
    & GO:0004867 & 0.742 & 0.884 & 0.807 & \textbf{0.972} & 0.568 & \red{0.175}& 0.955 & 0.985 & \textbf{0.992} \\
    & GO:0005096 & 0.904 & \textbf{0.907} & 0.750 & 0.847 & 0.517 & \red{0.359}& \textbf{0.962} & 0.938 & 0.735 \\
    \midrule 
    \multirow{7}{*}{BP} 
    & GO:0000381 & 0.973 & \textbf{0.988} & 0.957 & 0.855 & 0.751 & \red{0.423}& 0.987 & 0.906 & \textbf{0.992} \\
    & GO:0032729 & 0.895 & 0.895 & 0.948 & 0.870 & \textbf{0.982} & \red{0.866}& 0.893 & 0.932 & \textbf{0.996} \\
    & GO:0032755 & 0.895 & 0.889 & 0.957 & 0.719 & \textbf{0.966} & \red{0.754}& 0.939 & 0.884 & \textbf{0.979} \\
    & GO:0032760 & 0.900 & 0.883 & \textbf{0.930} & 0.861 & 0.803 & \red{0.691}& \textbf{0.950} & 0.925 & 0.821 \\
    & GO:0046330 & 0.949 & \textbf{0.960} & 0.923 & 0.855 & 0.866 & \red{0.496}& \textbf{0.966} & 0.904 & 0.888 \\
    & GO:0051897 & 0.893 & \textbf{0.942} & 0.913 & 0.772 & 0.929 & \red{0.683}& \textbf{0.962} & 0.888 & 0.957 \\
    & GO:0120162 & 0.734 & 0.779 & \textbf{0.811} & 0.637 & 0.491 & \red{0.600}& 0.817 & 0.738 & \textbf{0.957} \\
    \midrule 
    \multirow{4}{*}{CC} 
    & GO:0005762 & \textbf{0.998} & \textbf{0.998} & 0.991 & 0.889 & 0.600 & \red{0.090} & \textbf{0.999} & 0.874 & 0.995 \\
    & GO:0022625 & \textbf{0.989} & 0.937 & 0.950 & 0.898 & 0.487 & \red{0.397}& \textbf{0.995} & 0.893 & 0.991 \\
    & GO:0042788 & 0.940 & \textbf{0.963} & 0.930 & 0.858 & 0.532 & \red{0.520}& \textbf{0.990} & 0.889 & \textbf{0.990} \\
    & GO:1904813 & 0.675 & 0.806 & \textbf{0.903} & 0.653 & 0.658 & \red{0.724}& 0.895 & 0.792 & \textbf{0.955} \\
    \bottomrule
  \end{tabular}}
  \renewcommand{\arraystretch}{1}
\end{table*}

For zero-shot evaluation, we compared \model with DeepGOZero, DeepGO-SE, \red{and TransFew}, which have zero-shot capabilities, using the zero-shot dataset, where selected GO terms were excluded from all protein–term associations. To assess the contributions of different GO information sources, we ablated the structural and textual components of our GO embedding module, denoted STAR$_S$ (structural only, using anc2vec embeddings) and STAR$_T$ (textual-only, using SBERT-BioBERT embeddings), and STAR$_{ST}$ model combining both representations. All models, including baselines, were trained on the zero-shot dataset and evaluated on the full test set, with term-specific AUCs reported for held-out GO terms in Table~\ref{tab:zero-shot-comparison}.

Across the 16 held-out GO terms, STAR-GO variants collectively achieve the highest zero-shot AUCs in 13 cases, consistently outperforming DeepGOZero, DeepGO-SE \red{and TransFew}.
Performance varies by ontology and GO embeddings:
STAR$_T$ achieves the best results on most Molecular Function and Biological Process terms (e.g., GO:0001228 = 0.949, GO:0046330 = 0.960), highlighting the strength of semantic information captured from textual definitions.
STAR$_S$ performs best on a few terms (e.g., GO:0032760 = 0.930, GO:1904813 = 0.903) but performs poorly for MF terms.
The combined STAR$_{ST}$ model maintains stable and competitive performance across all ontologies, closely matching the textual variant.

An independent subsumption-prediction evaluation of the GO embeddings further contextualized these results (see Supplementary Table S3).
The anc2vec performed the best in recovering hierarchical parent–child relations, confirming its strong topological representation of GO.
However, anc2vec embeddings generalized less effectively to unseen terms, particularly for MF terms, resulting in the lower zero-shot AUCs of STAR$_S$.
Conversely, SBERT-BioBERT embeddings performed worse in subsumption prediction but generalized better to unseen terms, indicating that semantic representations transfer more effectively to novel ontology concepts.
The integrated STAR$_{ST}$ model reflects this complementarity, achieving balanced performance across ontologies by leveraging both semantic and structural information. While STAR$_S$ requires retraining when new GO terms are added due to the fixed anc2vec embeddings, STAR$_T$ and STAR$_{ST}$ can generate embeddings for novel terms directly from textual definitions, enabling continuous adaptation to ontology updates without retraining. \red{A seed sensitivity analysis over five random seeds confirmed the stability of these zero-shot results, with low standard deviations across all held-out terms (see Supplementary Table S5 and Table S6).}

\subsection{Hierarchical ordering and ontology-informed embeddings improve performance}

\begin{table*}[!htp]
  \tabcolsep=0pt
  \small
  \caption{Ablation study of \model across different ontologies and three aspects: (1) GO embeddings: None (learned from scratch), Structural (anc2vec), Textual (SBERT-BioBERT), or Combined (proposed integration of structural and textual embeddings);
    (2) Architecture: decoder with causal attention, encoder with bidirectional
    attention, or MLP using concatenated embeddings
    of the mean-pooled
    protein embeddings and GO embeddings; (3) GO term ordering:
    whether terms are
    hierarchically ordered in the decoder (\checkmark).
    Results are shown for
    Biological Process (BP), Cellular Component (CC), and Molecular
    Function (MF) predictions
    using multiple metrics. (-) indicates experiments not completed due to
  GPU memory constraints. The last configuration corresponds to the proposed model, \model.}
  \label{tab:ablation}
\begin{tabular*}{\textwidth}{@{\extracolsep{\fill}}lllcccccccccccc@{\extracolsep{\fill}}}
  \toprule
  \multicolumn{3}{l}{Model Configurations} &
  \multicolumn{3}{c}{Macro AUPR} &
  \multicolumn{3}{c}{Micro AUPR} &
  \multicolumn{3}{c}{AUC} &
  \multicolumn{3}{c}{Fmax} \\
  \midrule
  GO Embeddings & Architecture & GO Ordering &
  BP & CC & MF &
  BP & CC & MF &
  BP & CC & MF &
  BP & CC & MF \\
  \midrule
  None        & Decoder  & \checkmark   & 0.230 & 0.378 & 0.583 & 0.315 & 0.457 & 0.634 & 0.985 & 0.988 & 0.992 & 0.548 & 0.658 & 0.691 \\
  Structural  & Decoder  & \checkmark   & 0.267 & 0.366 & 0.608 & 0.322 & 0.449 & 0.660 & 0.983 & 0.989 & 0.994 & \textbf{0.553} & 0.653 & 0.700 \\
  Textual     & Decoder  & \checkmark   & 0.195 & 0.296 & 0.388 & 0.250 & 0.344 & 0.346 & 0.975 & 0.980 & 0.981 & 0.438 & 0.575 & 0.393 \\
  Combined    & MLP      & x            & 0.277 & 0.351 & 0.567 & 0.343 & 0.453 & 0.622 & \textbf{0.990} & 0.992 & \textbf{0.995} & 0.534 & 0.662 & 0.654 \\
  Combined    & Encoder  & x            & -     & 0.340 & 0.577 & -     & \textbf{0.463} & 0.633 & -     & \textbf{0.990} & \textbf{0.995} & - & \textbf{0.667} & 0.670 \\
  Combined    & Decoder  & x            & 0.281 & 0.361 & 0.610 & 0.340 & 0.443 & 0.650 & 0.985 & 0.989 & 0.993 & 0.542 & 0.647 & 0.707 \\

  Combined    & Decoder  & \checkmark   & \textbf{0.288}   & \textbf{0.379}   & \textbf{0.620}   & \textbf{0.351}   & 0.455            & \textbf{0.675}   & 0.989            & 0.988            & \textbf{0.995}   & 0.548            & 0.659            & \textbf{0.719} \\
  \bottomrule
\end{tabular*}
\end{table*}

We ablated \model to quantify the contributions of GO embedding design, integration architecture, and hierarchical ordering of GO terms. Table~\ref{tab:ablation} shows that the best performance was achieved by the decoder-based model using our integrated GO embedding module with hierarchically ordered GO terms. This variant shows  strength in Macro AUPR scores, achieving $0.288$,
$0.379$, and $0.620$ for BP, CC, and MF, respectively, and achieves
the highest $F_{\text{max}}$
score ($0.719$) for MF subontology. 

Removing either the textual or structural component from the GO embeddings results in performance degradation, most notably for BP, where Macro AUPR drops from $0.288$ to $0.267$ (structural-only) or $0.195$ (textual-only).
The textual-only variant performed substantially worse here than in the zero-shot setting, where it achieved the highest AUCs.
This discrepancy suggests that textual embeddings (SBERT-BioBERT) are effective for transferring to unseen terms but less so when trained end-to-end, whereas structural embeddings (anc2vec) provide a stronger inductive bias and stability when full supervision is available. \model performed the best across ontologies leveraging textual and structural information.

Replacing hierarchical ordering with a flat ordering results in an overall decrease in performance, confirming that the hierarchical ordering contributes a useful inductive bias by enforcing the GO structure.
Training GO embeddings from scratch (the ``None" configuration in Table~\ref{tab:ablation}, without any structural or textual pretrained GO embeddings) performs significantly worse, demonstrating the importance of pretrained GO embeddings for generalization.

We further compared different strategies for integrating GO embeddings with protein representations. In the decoder-based configuration, GO embeddings attend to each other with a causal mask such that GO terms are processed in hierarchical or GO ID-based order. The encoder-based configuration, which attends GO term sequence embeddings bidirectionally without ordered decoding, achieved competitive AUCs in CC and MF (0.990 and 0.995) and the highest CC $F_{\text{max}}$ ($0.667$), but could not be evaluated on BP due to memory constraints. The MLP integration, which pairs mean-pooled protein embeddings with individual GO term embeddings, produced similarly high AUCs (up to 0.995) but underperformed in $F_{\text{max}}$, Macro AUPR, and Micro AUPR compared to the decoder variant with GO hierarchical ordering. \red{Additionally, we evaluated ESM-1 and ESM-2 protein embeddings and found that ProtT5 performed better for our model (see Supplementary Table S4).}

\subsection{\red{\model discovers key residues in the zero-shot setting}}
\red{
The decoder cross-attention weights indicate protein residues the model considers relevant to each GO term.
To aggregate the signals across heads and layers, we applied recursive averaging to cross-attention matrices, inspired by attention rollout~\citep{abnar2020quantifying}.
Let $\bar{\mathbf{A}}^{(l)} = \frac{1}{H}\sum_{h=1}^{H}\mathbf{A}^{(l)}_h \in \mathbb{R}^{T \times L}$ denote the head-averaged cross-attention at decoder layer $l$. The rollout is computed recursively as:
\begin{align}
\mathbf{R}^{(1)} &= \bar{\mathbf{A}}^{(1)} \nonumber \\ \mathbf{R}^{(l)} &= \tfrac{1}{2}\bigl(\bar{\mathbf{A}}^{(l)} + \mathbf{R}^{(l-1)}\bigr), \quad l = 2,\ldots, N_{\text{dec}},
\label{eq:rollout}
\end{align}
equally weighting the current layer's attention and the accumulated score. Row $t$ of the final rollout $\mathbf{R} = \mathbf{R}^{(N_{\text{dec}})} \in \mathbb{R}^{T \times L}$ yields the per-residue attention profile $\mathbf{r}_t \in \mathbb{R}^L$ for GO term $t$.
}
\red{We applied this procedure to \model in the zero-shot setting to determine whether cross-attention captures biologically meaningful signals for a function the model has never seen.
We selected two held-out Molecular Function terms related to DNA-binding transcription factor activity: GO:0001228 (\textit{transcription activator}, zero-shot AUC = 0.938) and GO:0001227 (\textit{transcription repressor}, zero-shot AUC = 0.891). We computed attention rollout scores for three PDB chains: (GO:0001228)
4AWL chain~B (NF-YB),
2F8X chain~C (RBPJ/CSL), and (GO:0001227) 2HDC chain~A (FOXD3).
As ground truth for DNA-binding residues, we obtained experimentally determined DNA-contact sites from the BioLiP database~\citep{biolip}}.

\begin{figure*}[!htp]
  \centering
  \begin{minipage}[t]{0.48\textwidth}
    \centering
    \textbf{(a)}\\[2pt]
    \includegraphics[width=\linewidth]{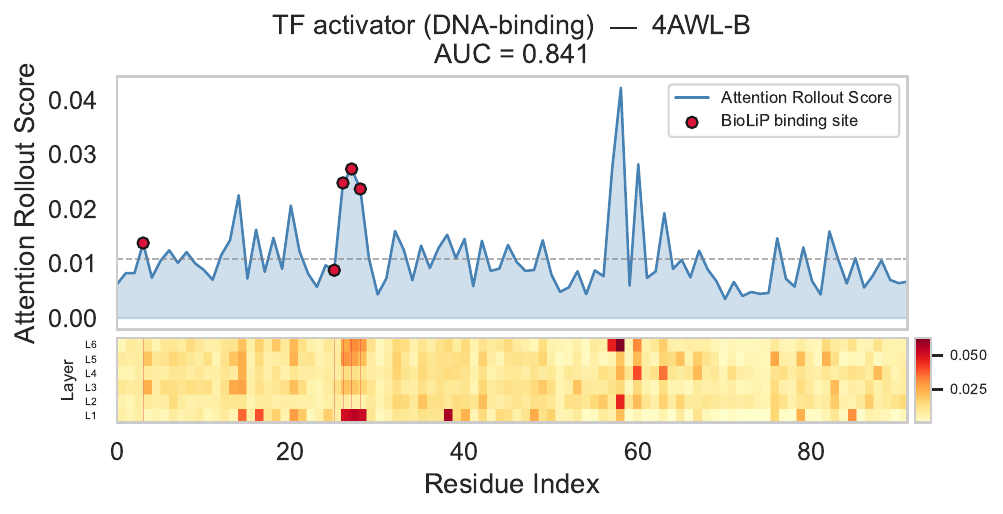}
  \end{minipage}%
  \hfill
  \begin{minipage}[t]{0.21\textwidth}
    \centering
    \textbf{(b)}\;\\\textsf{4AWL-B}\\[2pt]
    \includegraphics[width=\linewidth]{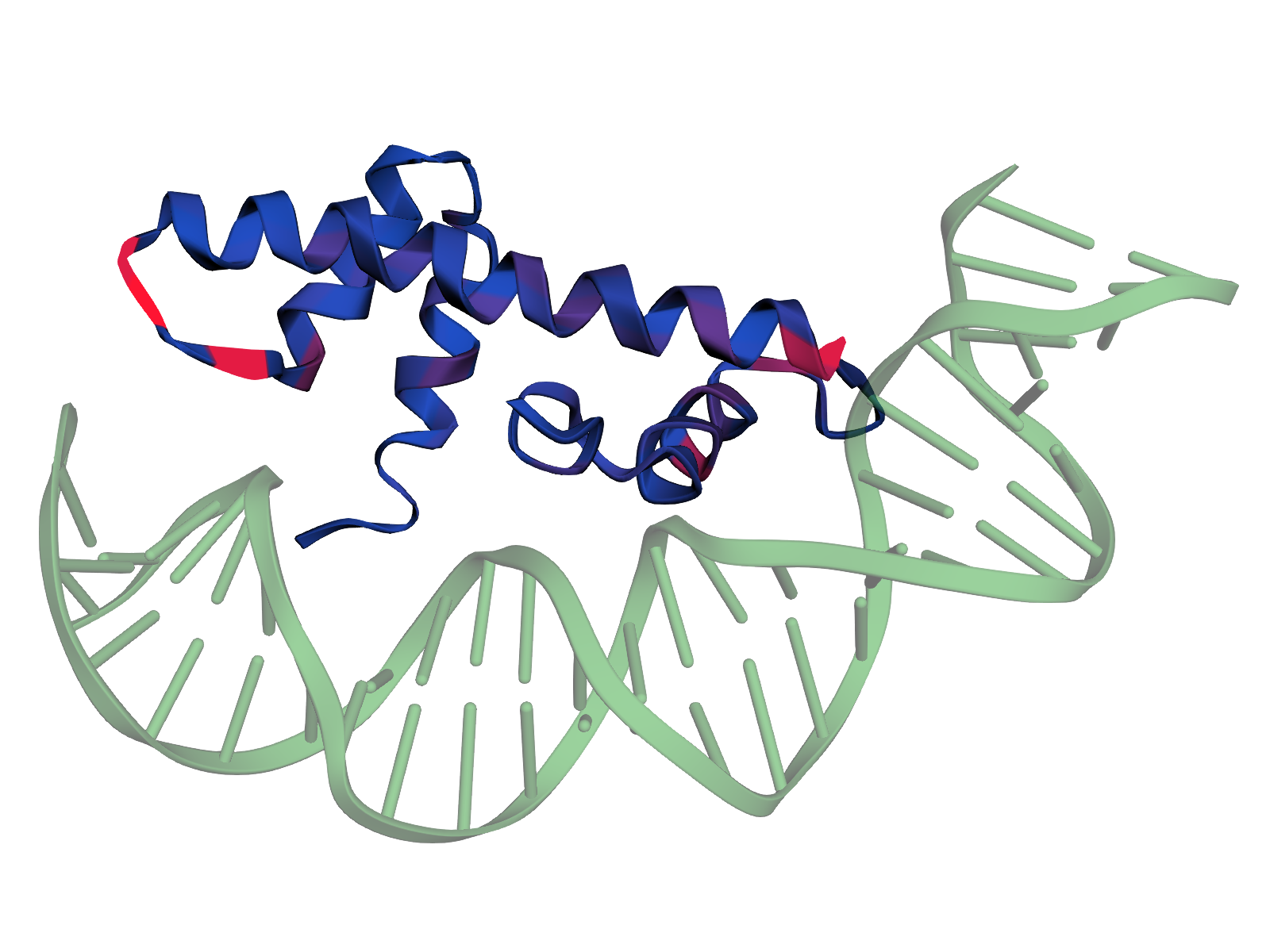}
  \end{minipage}%
  \hfill
  \begin{minipage}[t]{0.29\textwidth}
    \centering
    \textbf{(c)}\\[2pt]
    \includegraphics[width=\linewidth]{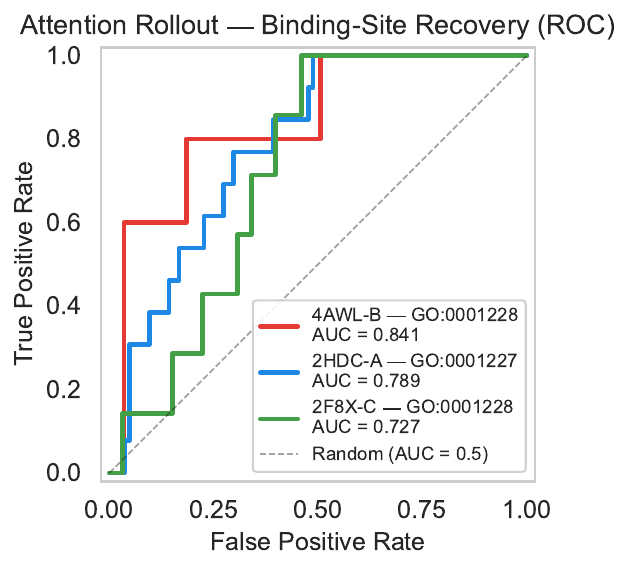}
  \end{minipage}
  \caption{\red{Interpretability analysis of STAR-GO's zero-shot cross-attention for two held-out DNA-binding transcription factor terms: GO:0001228 (activator) and GO:0001227 (repressor).
  \textbf{(a)}~Attention rollout and binding-site residues for 4AWL-B. Above: per-residue attention rollout scores with DNA-contact sites (red). Below: heatmap of per-residue attention averaged over heads for each decoder layer.
  \textbf{(b)}~3-D structure of 4AWL-B coloured by attention rollout score (blue\,$\to$\,red indicates increasing attention), with DNA shown in green.
  \textbf{(c)}~ROC curves evaluating agreement between attention-derived residue scores and BioLiP binding-site annotations.}}
  \label{fig:interpretability}
\end{figure*}

\red{Figure~\ref{fig:interpretability}a shows the per-residue attention rollout profile for 4AWL-B, with head-averaged attention at each decoder layer as a heatmap and the aggregated rollout score plotted above.
The BioLiP-annotated DNA-contact residues (red dots) coincide with regions of elevated attention, particularly around residues 25-30, corresponding to the NF-YB DNA-contact region, with a strong signal in the early decoder layers.
The 3D structure (Figure~\ref{fig:interpretability}b) confirms that residues receiving high rollout scores (red) are spatially proximal to the bound DNA (green).
Figure~\ref{fig:interpretability}c reports ROC curves for all three chains: 4AWL-B achieves an AUROC of 0.841, 2HDC-A 0.789, and 2F8X-C 0.727, all substantially above the random baseline.}

\red{Critically, both GO:0001228 and GO:0001227 were excluded from all protein-term associations during training; \model has never observed any protein annotated with either term.
\model's elevated attention to experimentally validated DNA-contact residues arises solely from the information in the term's embedding and its alignment with sequence features learned through supervision on other GO terms.
This demonstrates that the decoder learns a generalizable mapping between GO term semantics and residue-level features that extends beyond the training vocabulary.}

\section{Discussion}


This work demonstrates that integrating semantic and structural information yields a richer representation of GO terms for protein function prediction. The proposed GO embedding module, integrated within an encoder-decoder framework and combined with hierarchical decoding, enhances zero-shot prediction performance while maintaining competitive protein-centric and improved term-centric results.
Moreover, because our GO embedding method requires only term descriptions at inference, it can be applied across GO releases without retraining.

The ablation results provide additional insight into the interaction between semantic and structural information. The combined model does not consistently outperform the single-modality variants. Both the text-only and structure-only models achieve higher AUCs for certain terms, suggesting that each information source captures complementary yet occasionally conflicting aspects of functional similarity. This variability indicates that the current fusion mechanism may sometimes overemphasize one modality, depending on the characteristics of individual terms. Future work will develop adaptive fusion strategies that dynamically balance semantic and structural contributions, improving robustness across GO terms. Our implementation employs specific pretrained language models for both protein and GO embeddings. However, the framework can incorporate alternative protein representations or additional modalities, such as structure-based or protein–protein interaction embeddings. Future work will also explore the fusion of complementary representations to further enhance the performance.

\section{Conclusion}

We presented \model, a Transformer-based framework that jointly embeds the semantic and structural characteristics of GO terms to enhance zero-shot protein function prediction. The proposed GO embedding module refines language-model-derived representations through structural regularization, preserving both textual semantics and hierarchical relationships of GO terms. This design enables compatibility with evolving GO releases without retraining, while hierarchical decoding provides an inductive bias for propagating information from general to specific terms. Our results demonstrate that \model achieves strong zero-shot generalization and competitive performance compared to state-of-the-art methods. Our findings highlight that semantic and structural representations capture complementary aspects of functional similarity: textual embeddings provide adaptability and compatibility across GO releases, while structural information encodes relational context within the ontology. Future work will focus on developing adaptive fusion mechanisms and extending \model to include additional modalities for both proteins and GO terms. This study represents an important step toward generalizable function prediction models that are resilient to the continual evolution of biological ontologies.

\section*{Competing interests}
No competing interest is declared.

\section*{Author contributions statement}
M.E.A., G.U., A.Ö. and İ.M.B. conceived and designed the study.
M.E.A. implemented the algorithms and conducted the experiments with
guidance from G.U. M.E.A. wrote the initial version of the
manuscript. G.U., A.Ö. and İ.M.B. analysed the results and reviewed
the manuscript.

\section*{Acknowledgments}
This work is supported by ERC grant (LifeLU, 101089287). Views and
opinions expressed are however those of the author(s) only and do not
necessarily reflect those of the European Union or the European
Research Council Executive Agency. Neither the European Union nor the
granting authority can be held responsible for them.

\bibliographystyle{abbrvnat}
\bibliography{references/proteinfunc,references/ontology,references/owl2vecrefs,references/pfresgorefs}

\appendix
\section{Supplementary Materials}

\renewcommand{\thetable}{S\arabic{table}}
\subsection*{Dataset statistics}

Table~\ref{tab:dataset-distribution} summarizes the number of protein sequences and annotated GO terms in the training, validation, and test splits used for model development. Table~\ref{tab:dataset-distribution-uniprot-swissprot-slim} presents the corresponding statistics for the dataset employed in zero-shot evaluation, including per-subontology protein and term counts.

\begin{table}[!htp]
  \centering
  \caption{Dataset statistics for training, validation
    and test splits. The number of protein sequences in each split and the number of GO
  terms with annotations per ontology.}
  \label{tab:dataset-distribution}
  \begin{tabular}{lcccc}
    \toprule
    Subontology & Train  & Test  & Validation & Num. terms \\
    \hline
    MF          & 29,902 & 3,416 & 3,323      & 489        \\
    BP          & 29,902 & 3,416 & 3,323      & 1,943      \\
    CC          & 29,902 & 3,416 & 3,323      & 320        \\
    \hline
  \end{tabular}
\end{table}

\begin{table}[!htp]
  \centering
  \caption{Statistics for UniProtKB\textendash SwissProt 2021\_4 dataset used in zero-shot evaluation. Per-subontology protein counts in each split and term count is shown.}
  \label{tab:dataset-distribution-uniprot-swissprot-slim}
  \begin{tabular}{lcccc}
    \toprule
    Subontology & Terms & Training & Validation & Testing \\
    \hline
    MF & 6,868 & 34,716 & 3,851 & 4,712 \\
    BP & 21,381 & 47,733 & 5,552 & 5,444 \\
    CC & 2,832 & 48,318 & 4,970 & 5,969 \\
    \hline
  \end{tabular}
\end{table}

\section*{GO Embedding Module}

\begin{figure}[!htp]
    \centering
    \includegraphics[width=\linewidth]{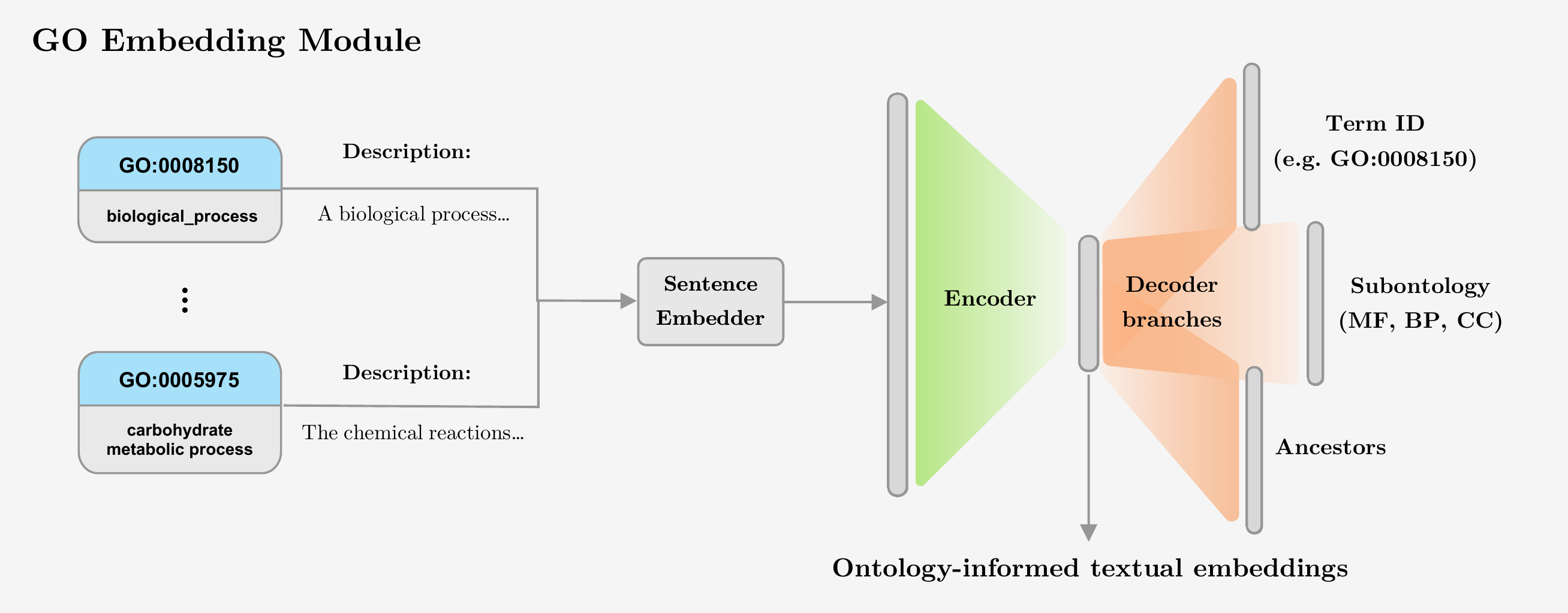}
    \caption{Overview of the GO embedding module. Sentence embeddings obtained from SBERT-BioBERT are projected into a latent space through a trainable autoencoder. The encoder maps GO term definitions into a semantic-structural representation, while three decoder branches jointly reconstruct (i) ancestor terms, (ii) subontology, and (iii) term identity. Multi-task supervision aligns the semantic embeddings of related GO terms with their hierarchical relationships in the ontology.}
    \label{fig:goembedder}
\end{figure}

Figure \ref{fig:goembedder} summarizes the GO embedding framework described in Section~\textit{GO Embeddings}, highlighting the integration of semantic (text-based) and structural (ontology-based) information via a projection autoencoder trained with multi-task objectives.

\section*{\red{Transformer Module}}

\red{The encoder and decoder of \model follow the standard pre-norm Transformer architecture~\citep{vaswaniAttentionAllYou2023} with multi-head attention (MHA), position-wise feed-forward networks (FFN), and layer normalization (LN). Below we provide the full equations for completeness.}

\subsection*{\red{Encoder Self-Attention}}

\red{Let $\mathbf{H}^{(l)}_{\text{enc}}\in\mathbb{R}^{L\times d}$ denote the hidden representation of the protein sequence at the $l$-th encoder layer, where $\mathbf{H}^{(0)}_{\text{enc}}$ is the projected input (see main text). Each layer updates these representations as follows:}
\begin{align}
\hat{\mathbf{H}}^{(l)}_{\text{enc}} &=
\mathrm{LN}\!\left(
\mathbf{H}^{(l-1)}_{\text{enc}}+
\mathrm{MHA}\!\left(
\mathbf{H}^{(l-1)}_{\text{enc}},
\mathbf{H}^{(l-1)}_{\text{enc}};
\tilde{\mathbf{M}}_{\text{seq}}
\right) \right),  \\
\mathbf{H}^{(l)}_{\text{enc}} &=
\mathrm{LN}\!\left(
\hat{\mathbf{H}}^{(l)}_{\text{enc}}+\mathrm{FFN}\!\left(\hat{\mathbf{H}}^{(l)}_{\text{enc}}\right)
\right)
\end{align}
\red{where $\mathrm{MHA}$ denotes multi-head attention, $\mathrm{FFN}$ is a position-wise feed-forward network, $\mathrm{LN}$ denotes layer normalization, and $\tilde{\mathbf{M}}_{\text{seq}}$ is an additive mask for sequence padding. After $N_{\text{enc}}$ layers, the encoder outputs refined residue embeddings $\mathbf{H}^{(N_{\text{enc}})}_{\text{enc}}$.}

\subsection*{\red{Decoder Self-Attention and Cross-Attention}}

\red{Let $\mathbf{H}^{(l)}_{\text{dec}}\in\mathbb{R}^{T\times d}$ denote the hidden representation of the GO terms at decoder layer $l$, where $\mathbf{H}^{(0)}_{\text{dec}}$ is the projected GO embedding input (see main text). Each decoder layer is computed as:}

\begin{align}
\textbf{Self-attention:}\;
\hat{\mathbf{H}}^{(l)}_{\text{dec}} &=
\mathrm{MHA}\!\left(
\mathbf{H}^{(l-1)}_{\text{dec}},
\mathbf{H}^{(l-1)}_{\text{dec}};
\tilde{\mathbf{M}}_{\text{dec}}
\right), \\
\mathbf{S}^{(l)}_{\text{dec}} &=
\mathrm{LN}\!\left(
\mathbf{H}^{(l-1)}_{\text{dec}}+\hat{\mathbf{H}}^{(l)}_{\text{dec}}
\right), \\[4pt]
\textbf{Cross-attention:}\;
\tilde{\mathbf{H}}^{(l)}_{\text{dec}} &=
\mathrm{MHA}\!\left(
\mathbf{S}^{(l)}_{\text{dec}},
\mathbf{H}^{(N_{\text{enc}})}_{\text{enc}};
\mathbf{0}
\right), \\
\mathbf{C}^{(l)}_{\text{dec}} &=
\mathrm{LN}\!\left(
\mathbf{S}^{(l)}_{\text{dec}}+\tilde{\mathbf{H}}^{(l)}_{\text{dec}}
\right), \\[4pt]
\textbf{Feed-forward:}\;
\mathbf{H}^{(l)}_{\text{dec}} &=
\mathrm{LN}\!\left(
\mathbf{C}^{(l)}_{\text{dec}}+\mathrm{FFN}\!\left(\mathbf{C}^{(l)}_{\text{dec}}\right)
\right).
\end{align}

\red{The decoder's self-attention mask $\tilde{\mathbf{M}}_{\text{dec}}$ is a lower-triangular matrix that respects the topological ordering of GO terms, ensuring that each term attends only to its ancestors and preceding terms in the hierarchy. Cross-attention uses no mask ($\mathbf{0}$), allowing each GO term to attend to all encoded protein residues.}

\subsection*{\red{Prediction Head and Training Objective}}

\red{For each GO term $t$, the final decoder output is mapped through a two-layer feed-forward projection:}
\begin{align}
\mathbf{u}_t &= \mathrm{GELU}\!\left(
\mathbf{H}^{(N_{\text{dec}})}_{\text{dec}}[t,:]\mathbf{W}_1+\mathbf{b}_1
\right), \\
z_t &= \mathbf{u}_t\mathbf{W}_2 + b_2, \quad
\hat{y}_t = \sigma(z_t)\in(0,1).
\end{align}
\red{where $\hat{y}_t$ denotes the predicted probability that the protein is associated with GO term $t$, and $\sigma(\cdot)$ is the sigmoid activation function.}

\red{Given binary ground-truth $\mathbf{y}\in\{0,1\}^{T}$, where $y_t = 1$ if the protein is annotated with GO term $t$ and $0$ otherwise, the model is trained using binary cross-entropy:}
\begin{equation}
\mathcal{L}_{\text{BCE}} =
\frac{1}{T}\sum_{t=1}^{T}
\Big(
\log(1+\exp(z_t)) - y_t z_t
\Big).
\end{equation}

\section*{Evaluation Metrics}
We evaluate performance using the standard metrics of the Critical Assessment of Functional Annotation (CAFA) challenges\citep{zhou2019cafa}
including macro-averaged Area Under the Precision-Recall Curve
(AUPR), $F_{\text{max}}$ score,
and micro-averaged Area Under the ROC Curve (AUC). Before evaluation, all predictions and labels are propagated up the Gene Ontology hierarchy according to the true path rule.

\begin{itemize}
  \item \textbf{Protein-centric $\mathbf{F_{max}}$:} This is the protein-centric
    maximum for the $F_{1}$ measure. For a threshold $t$, define precision
    $P_{t}= \frac{TP_{t}}{TP_{t}+ FP_{t}}$ and recall
    $R_{t}= \frac{TP_{t}}{TP_{t}+ FN_{t}}$, where $TP_{t}$, $FP_{t}$, and $F
    N_{t}$ represents true positives, false positives, and false negatives at
    threshold $t$. The F-measure at threshold $t$ is calculated as
    $F_{t}= \frac{2 \cdot P_{t}\cdot R_{t}}{P_{t}+ R_{t}}$. $F_{max}$ is defined
    as:
    \begin{equation}
      F_{max}= \max_{t \in [0,1]}F_{t}
    \end{equation}
    When evaluating protein function prediction, the protein-centric $\mathbf{F_{max}}$ accounts for
    the GO term hierarchy by propagating predictions along the
    ontology structure.

  \item \textbf{Macro AUPR:} For each GO term $g$ in the set of all terms $G$, an individual AUPR value is calculated as:
    \begin{equation}
      \text{AUPR}_{g}= \int_0^1 P_{g}(r)dr
    \end{equation}
    where $P_{g}(r)$ is the precision at recall level $r$ for term
    $g$. The Macro
    AUPR is then:
    \begin{equation}
      \text{Macro AUPR}= \frac{1}{|G|}\sum_{g \in G}\text{AUPR}_{g}
    \end{equation}
    This gives equal weight to each term regardless of its prevalence in the
    dataset.

  \item \textbf{AUC (Area Under the ROC Curve):} For ROC calculation, the true
    positive rate $\text{TPR}= \frac{TP}{TP+FN}$ is plotted against the false
    positive rate $\text{FPR}= \frac{FP}{FP+TN}$ at various thresholds. The micro-averaged AUC is calculated by first flattening all predictions and
    true labels across all protein-term pairs into a single vector,
    then calculating
    \begin{equation}
      \text{AUC}= \int_{0}^{1}\text{TPR}(\text{FPR}^{-1}(f)) \, df
    \end{equation}
    where $\text{FPR}^{-1}(f)$ is the threshold that gives an FPR of $f$.
\end{itemize}

The metrics above, calculated as defined in CAFA guidelines, provide a comprehensive
view of model performance, covering both protein-level accuracy
($F_{\text{max}}$, AUC) and
term-specific prediction quality (Macro AUPR).

\section*{Baselines}

We evaluated \model against five baseline methods selected for their methodological relevance, zero-shot capabilities, or state-of-the-art performance.

\textbf{Methods with similar architectures}. PFresGO~\citep{panPFresGOAttentionMechanismbased2023} combines ProtT5 protein representations with anc2vec GO embeddings through cross-attention in an auto-encoder framework with non-feedforward residual attention blocks. Unlike \model's encoder-decoder architecture with semantic GO representations, PFresGO relies solely on graph-derived hierarchical information. TALE~\citep{caoTALETransformerbasedProtein2021} employs a Transformer encoder to learn joint protein-GO embeddings, using convolution with softmaxed similarity scores for prediction.

\textbf{Zero-shot capable methods}. DeepGOZero~\citep{kulmanov2022deepgozero} pioneered zero-shot protein function prediction by representing GO terms as geometric n-balls that satisfy formal Description Logic axioms, operating on InterPro domain features rather than sequence embeddings. DeepGO-SE~\citep{kulmanovProteinFunctionPrediction2024} extends this geometric approach through approximate semantic entailment across multiple models while incorporating ESM2 protein language models. \red{More recently, TransFew~\citep{transfew} follows a method similar to STAR-GO by combining BioBert embeddings of GO definitions with GO hierarchy through a GCN, and cross-attends them with refined ESM embeddings.} 

\textbf{Structure-based}. DeepFRI~\citep{gligorijevicStructurebasedProteinFunction2021} is a structure-based function predictor, using graph convolutional networks on residue contact maps to capture spatial functional determinants.

\section*{Implementation Details}
The models were trained and evaluated using the PyTorch 2.5.1
\citep{paszke2017automatic},
HuggingFace Transformers 4.47.0
\citep{DBLP:journals/corr/abs-1910-03771}, and PyTorch
Lightning 2.4.0\citep{Falcon_PyTorch_Lightning_2019} libraries. Our
model implementation
utilized HuggingFace's BERT encoder and decoder layers, replacing the trained
embeddings module with the frozen embeddings from our method. For the
encoder variant,
we utilized a fully enabled attention mask in the self-attention
layer, effectively
making it an encoder.

Experiment tracking and hyperparameter tuning were performed on the
Weights\&Biases
platform \citep{wandb}. Hyperparameter tuning was utilized to optimize
the model's
performance and experiments were performed with multiple variants of the model. In
particular, we experimented with the fixed ordering of Gene Ontology
terms in the
decoder input, testing the variations of: ordered, unordered GO terms with a
causal mask and unordered GO terms without a causal mask. We also evaluated an ablation variant in which protein and GO term embeddings were concatenated and passed through an MLP prediction head.  We utilized PyTorch Lightning's
Learning Rate Finder to find the optimal learning rate for each model
architecture.

For the GO embedding module fine-tuning, we adapted the Anc2vec training procedure to incorporate our semantic embeddings using a multi-task loss formulation with an ExponentialDecay learning rate scheduler. The embedding dimension was set to $d=200$. Training proceeded for $100$ epochs with batch size $32$, using scheduler parameters \verb|initial_lr=0.001|, \verb|decay_rate=0.9|, and \verb|decay_steps=10000|. Loss weights were set to $\alpha_\text{anc}=1.0$, $\alpha_\text{sub}=0.5$, and $\alpha_\text{id}=0.3$ for the ancestor prediction, subontology classification, and term identity reconstruction tasks, respectively.

For \model, we use a hidden size $d=256$ in all Transformer blocks with $6$ encoder and $6$ decoder layers, each with $8$ attention heads and a feed-forward intermediate dimension of $1024$. Dropout is $0.1$ on hidden activations and $0.1$ on attention probabilities; activations use \textsc{GELU} and we set \texttt{layer\_norm\_eps} to $1!\times!10^{-12}$. GO embeddings are $200$-d for Anc2Vec and our learned GO embeddings, and $768$-d when using the SBERT-BioBERT textual encoder; protein residues are projected to the shared hidden size $d$ 
We train with AdamW (learning rate $6{\times}10^{-5}$ selected by a PyTorch Lightning learning-rate finder, weight decay $0.01$), without a learning-rate scheduler, for up to $100$ epochs with early stopping (patience $10$) on minimum validation loss. To standardize compute, we employ gradient accumulation together with Fully Sharded Data Parallel training, yielding an effective batch size of $32$ across runs. Mixed precision uses \texttt{bf16} (\texttt{bf16-mixed} in PyTorch). 

We adopted PFresGO's official evaluation code\footnote{\url{https://github.com/BioColLab/PFresGO}}
 for metric computation. All baseline results are taken from the PFresGO study, except for DeepGO-SE, which we retrained and evaluated for consistency. For zero-shot experiments, both DeepGOZero and DeepGO-SE were retrained under our ablation protocol.

\section*{Subsumption prediction with GO embeddings}

We compared our GO embedding module against several baseline GO embedding techniques on the subsumption prediction task (Table~\ref{tab:go-embedding}), including OWL2Vec*~\citep{chenOWL2VecEmbeddingOWL2021}, anc2Vec~\citep{ederaAnc2vecEmbeddingGene2022}, and GT2Vec~\citep{zhaoLearningRepresentationsGene2022}. To assess the contribution of semantic information, we also evaluated pretrained language models that encode GO term definitions without structural supervision. Specifically, we included two task-specific BERT models: \textit{BioBERT}~\citep{leeBioBERTPretrainedBiomedical2020}, pretrained on large-scale biomedical text, and \textit{SBERT-BioBERT}~\citep{reimersSentenceBERTSentenceEmbeddings2019a}, which fine-tunes BioBERT on sentence similarity tasks using the SentenceTransformers framework.

The evaluation was performed using the GO subsumption dataset and evaluation
metrics of OWL2Vec*~\citep{chenOWL2VecEmbeddingOWL2021}. The dataset
is a split of
all GO subsumption triples, i.e., \verb|is_a|, \verb|part_of|
relations between GO
terms. We evaluated the \verb|go-basic| and \verb|go| editions of the
GO. The basic
edition contains \textit{part of}, \textit{is a}, \textit{regulates},
and \textit{negatively
regulates}. The full edition \verb|go| additionally contains \textit{has part}
and \textit{occurs in}. It is generally considered unsafe to propagate
annotations with the complete edition since the additional relations introduce
cycles to the graph~\citep{geneontologyconsortiumGeneOntologyKnowledgebase2023}.

For each embedding technique, we trained multiple models to predict
the likelihood of a subsumption relationship between any two GO terms
and selected the best-performing one. This model was then used for a
ranking task: for each child term in the test set, we ranked all
other GO terms as potential parents based on the predicted
likelihood. We evaluated these rankings using standard information
retrieval metrics. Hits@k measures the percentage of times a true
parent is found within the top k ranked candidates; we report for
k=1, 5, and 10. Mean Reciprocal Rank (MRR) evaluates the average
inverse rank of the first correct parent, providing a single score
for overall ranking quality. Higher values for all metrics indicate
better performance.

As shown in Table~\ref{tab:go-embedding}, anc2vec achieved the highest accuracy in recovering subsumption relations, consistent with its explicit optimization for structural hierarchy. Sentence-based embeddings performed worse overall, though fine-tuned semantic embeddings (SBERT-BioBERT) provided a substantial improvement over non-fine-tuned BioBERT, demonstrating the effectiveness of fine-tuning the SBERT-BioBERT embeddings.
Our GO embedding module, which integrates semantic embeddings with structural reconstruction objectives, performed comparably to OWL2Vec* across metrics and surpassed purely semantic models. Despite anc2vec's stronger subsumption scores, our ablation study (main text, Table 3) showed that these embeddings generalized less effectively to downstream protein function prediction. In contrast, our module's semantic–structural integration improved transfer to function prediction, confirming the value of including semantic information alongside hierarchical structure.

\begin{table}[H]
  \centering
  \caption{Performance of GO embedding techniques. anc2vec and OWL2Vec* are
  the same as in their respective papers. BioBERT uses mean-pooled
  sentence embeddings of term descriptions. SBERT-BioBERT is a
  BioBERT model fine-tuned for sentence similarity tasks. Our GO embedding
  method fine-tunes SBERT-BioBERT further with structure recovery
  objectives. We evaluated on two editions of GO: full and basic.}
  \label{tab:go-embedding}

  \setlength{\tabcolsep}{0pt}

  \begin{tabular*}{\textwidth}{@{\extracolsep{\fill}}lllrrrr@{}}
    \toprule
    GO Edition & Method &  & Hits@1 & Hits@10 & Hits@5 & MRR \\
    \midrule
    \multirow[t]{5}{*}{go} &
    anc2vec       &  & \textbf{0.077} & \textbf{0.365} & \textbf{0.245} & \textbf{0.170} \\
    & Ours         &  & 0.063          & 0.333          & 0.220          & 0.149 \\
    & OWL2Vec*     &  & 0.071          & 0.343          & 0.230          & 0.158 \\
    & SBERT-BioBERT&  & 0.060          & 0.297          & 0.190          & 0.137 \\
    & BioBERT      &  & 0.008          & 0.112          & 0.058          & 0.047 \\
    \midrule
    \multirow[t]{6}{*}{go-basic} &
    anc2vec       &  & \textbf{0.092} & \textbf{0.414} & \textbf{0.290} & \textbf{0.196} \\
    & Ours         &  & 0.069          & 0.340          & 0.233          & 0.157 \\
    & OWL2Vec*     &  & 0.066          & 0.333          & 0.221          & 0.152 \\
    & SBERT-BioBERT&  & 0.054          & 0.292          & 0.188          & 0.131 \\
    & BioBERT      &  & 0.022          & 0.156          & 0.097          & 0.071 \\
    & GT2Vec       &  & 0.006          & 0.067          & 0.028          & 0.038 \\
    \bottomrule
  \end{tabular*}
\end{table}

\section*{\red{Impact of residue embeddings across different protein language models}}

\red{To assess the impact of the protein language model on downstream function prediction, we compared three residue-level embedding methods within STAR-GO's architecture: ESM-1b~\citep{rives2019biological}, ESM-2~\citep{linEvolutionaryscalePredictionAtomic2022}, and ProtT5~\citep{ProtTransUnderstandingLanguage}. All other model components, including the GO embedding module, decoder architecture, and training procedure, were held constant; only the frozen residue embeddings provided to the encoder were varied. Table~\ref{tab:plm-ablation} reports results under our standard supervised evaluation protocol across all three ontologies.}

\red{ProtT5 embeddings consistently outperform both ESM variants across most metrics. The improvement is most pronounced for Molecular Function, where ProtT5 achieves an $F_{\text{max}}$ of $0.719$ compared to $0.675$ for both ESM-1b and ESM-2, and a Macro AUPR of $0.620$ versus $0.574$--$0.576$. Similar gains are observed for Biological Process and Cellular Component in Macro AUPR and Micro AUPR. AUC scores are comparable across all three embeddings, indicating that the discriminative advantage of ProtT5 is most evident in precision-recall--based metrics where ranking quality among positive terms matters most. These results motivated our adoption of ProtT5 as the default protein encoder in STAR-GO.}

\begin{table}[H]
  \tabcolsep=0pt
  \small
  \caption{\red{Comparison of protein language model embeddings for protein function prediction. We evaluate three residue-level embedding methods: ESM1b, ESM2, and ProtT5. Results are reported across Biological Process (BP), Cellular Component (CC), and Molecular Function (MF) ontologies. ProtT5 embeddings (used in \model) consistently outperform ESM variants across most metrics, with particularly notable gains in Macro AUPR and Fmax.}}
  \label{tab:plm-ablation}
\begin{tabular*}{\textwidth}{@{\extracolsep{\fill}}lcccccccccccc@{\extracolsep{\fill}}}
  \toprule
  \multicolumn{1}{l}{Embeddings} &
  \multicolumn{3}{c}{Macro AUPR} &
  \multicolumn{3}{c}{Micro AUPR} &
  \multicolumn{3}{c}{AUC} &
  \multicolumn{3}{c}{Fmax} \\
  \midrule
  Residue Embeddings &
  BP & CC & MF &
  BP & CC & MF &
  BP & CC & MF &
  BP & CC & MF \\
  \midrule
  ESM1b  & 0.253            & 0.329            & 0.574            & 0.321            & 0.408            & 0.631            & 0.980            & \textbf{0.988}            & \textbf{0.995}   & 0.539            & 0.639            & 0.675 \\
  ESM2   & 0.268            & 0.349            & 0.576            & 0.321            & 0.406            & 0.628            & 0.975            & 0.985            & 0.994            & 0.544            & 0.640            & 0.675 \\
   ProtT5  & \textbf{0.288}   & \textbf{0.379}   & \textbf{0.620}   & \textbf{0.351}   & \textbf{0.455}            & \textbf{0.675}   & \textbf{0.989}            & \textbf{0.988}            & \textbf{0.995}   & \textbf{0.548}            & \textbf{0.659}            & \textbf{0.719} \\
  \midrule
\end{tabular*}
\end{table}

\section*{\red{Zero-shot experiment sensitivity analysis}}

\red{To assess the stability of STAR-GO's zero-shot predictions with respect to random initialization, we repeated the zero-shot evaluation across five independent training seeds, keeping all hyperparameters fixed. Tables~\ref{tab:zero-shot-bp} and~\ref{tab:zero-shot-cc-mf} report per-term AUC scores for each seed along with the mean and standard deviation.}

\red{Across all 16 held-out GO terms, the model exhibits low variance in most terms, with most standard deviations below $0.03$. Cellular Component terms show particularly stable predictions, exemplified by GO:0005762 (mean AUC $= 0.9961 \pm 0.0016$). 
Overall, the consistently low standard deviations confirm that the zero-shot generalization is robust and not an artifact of a particular random seed.}

\begin{table}[ht]
\centering
\caption{\red{Zero-shot protein function prediction AUC scores (Biological Process) across 5 seeds.}}
\label{tab:zero-shot-bp}
\resizebox{\textwidth}{!}{%
\begin{tabular}{lccccccc}
\toprule
 & \multicolumn{7}{c}{Biological Process} \\
\cmidrule(lr){2-8}
Seed & GO:0000381 & GO:0032729 & GO:0032755 & GO:0032760 & GO:0046330 & GO:0051897 & GO:0120162 \\
\midrule
1 & 0.9735 & 0.8936 & 0.9042 & 0.9064 & 0.9685 & 0.9073 & 0.7754 \\
2 & 0.9831 & 0.9201 & 0.9152 & 0.9630 & 0.9006 & 0.9034 & 0.8103 \\
3 & 0.9709 & 0.8891 & 0.9030 & 0.9254 & 0.9609 & 0.9347 & 0.8290 \\
4 & 0.9715 & 0.9058 & 0.9070 & 0.8935 & 0.9281 & 0.9223 & 0.8051 \\
5 & 0.9849 & 0.9106 & 0.9162 & 0.9132 & 0.9556 & 0.9341 & 0.8115 \\
\midrule
Mean $\pm$ Std & $0.9768 \pm 0.0067$ & $0.9038 \pm 0.0126$ & $0.9091 \pm 0.0062$ & $0.9203 \pm 0.0265$ & $0.9427 \pm 0.0281$ & $0.9203 \pm 0.0146$ & $0.8063 \pm 0.0195$ \\
\bottomrule
\end{tabular}}
\end{table}

\begin{table}[ht]
\centering
\caption{\red{Zero-shot protein function prediction AUC scores (Cellular Component \& Molecular Function) across 5 seeds.}}
\label{tab:zero-shot-cc-mf}
\resizebox{\textwidth}{!}{%
\begin{tabular}{lccccccccc}
\toprule
 & \multicolumn{4}{c}{Cellular Component} & \multicolumn{5}{c}{Molecular Function} \\
\cmidrule(lr){2-5} \cmidrule(lr){6-10}
Seed & GO:0005762 & GO:0022625 & GO:0042788 & GO:1904813 & GO:0001227 & GO:0001228 & GO:0003735 & GO:0004867 & GO:0005096 \\
\midrule
1 & 0.9941 & 0.9672 & 0.8809 & 0.7675 & 0.8956 & 0.9271 & 0.7650 & 0.6246 & 0.9306 \\
2 & 0.9948 & 0.9216 & 0.9056 & 0.7231 & 0.9270 & 0.9476 & 0.9219 & 0.8552 & 0.9367 \\
3 & 0.9973 & 0.9746 & 0.9625 & 0.7587 & 0.9476 & 0.9526 & 0.6933 & 0.9030 & 0.9243 \\
4 & 0.9975 & 0.9622 & 0.9254 & 0.6598 & 0.9387 & 0.9413 & 0.8459 & 0.7896 & 0.9405 \\
5 & 0.9969 & 0.8599 & 0.9041 & 0.7939 & 0.9253 & 0.9442 & 0.9375 & 0.7253 & 0.9184 \\
\midrule
Mean $\pm$ Std & $0.9961 \pm 0.0016$ & $0.9371 \pm 0.0478$ & $0.9157 \pm 0.0305$ & $0.7406 \pm 0.0518$ & $0.9269 \pm 0.0197$ & $0.9426 \pm 0.0096$ & $0.8327 \pm 0.1039$ & $0.7796 \pm 0.1096$ & $0.9301 \pm 0.0090$ \\
\bottomrule
\end{tabular}}
\end{table}

\end{document}